\documentclass[10pt,conference]{IEEEtran}

\newcommand\copyrighttext{%
  \footnotesize \textcopyright~2025 IEEE. Personal use of this material is permitted. Permission from IEEE must be obtained for all other uses, in any current or future media, including reprinting/republishing this material for advertising or promotional purposes, creating new collective works, for resale or redistribution to servers or lists, or reuse of any copyrighted component of this work in other works.}
\newcommand\copyrightnotice{%
\begin{tikzpicture}[remember picture,overlay]
\node[anchor=south,yshift=10pt] at (current page.south) {\fbox{\parbox{\dimexpr\textwidth-\fboxsep-\fboxrule\relax}{\copyrighttext}}};
\end{tikzpicture}%
}

\makeatletter
\newcommand{\linebreakand}{%
  \end{@IEEEauthorhalign}
  \hfill\mbox{}\par
  \mbox{}\hfill\begin{@IEEEauthorhalign}
}
\makeatother

\IEEEoverridecommandlockouts
\usepackage{cite}
\usepackage{amsmath,amssymb,amsfonts}
\usepackage{algorithmic}
\usepackage{graphicx}
\usepackage{textcomp}
\usepackage{xcolor}
\def\BibTeX{{\rm B\kern-.05em{\sc i\kern-.025em b}\kern-.08em
    T\kern-.1667em\lower.7ex\hbox{E}\kern-.125emX}}

\usepackage{enumitem}
\usepackage{booktabs}
\usepackage{url}
\usepackage[most]{tcolorbox}
\usepackage[caption=false,font=footnotesize]{subfig}
\usepackage{tabularx}
\usepackage[hidelinks]{hyperref}
\usepackage{cleveref}

\definecolor{promptBg}{HTML}{F7F8FA}
\definecolor{promptBorder}{HTML}{000000}
\definecolor{promptTitle}{HTML}{4A4A4A}
\definecolor{promptVar}{HTML}{000000}
\definecolor{promptLabel}{HTML}{000000}
\newcommand{\promptvar}[1]{{\normalfont\color{promptVar}[\textbf{#1}]}}
\newcommand{\promptlabel}[1]{{\normalfont\color{promptLabel}\textbf{#1:}}}
\newtcolorbox{promptbox}[1][]{%
  enhanced,
  colback=promptBg,
  colframe=promptBorder,
  boxrule=0.4pt,
  arc=3pt,
  left=2pt, right=2pt, top=10pt, bottom=2pt,
  fonttitle=\bfseries\footnotesize,
  coltitle=white,
  colbacktitle=promptTitle,
  attach boxed title to top left={xshift=7pt, yshift=-7pt},
  boxed title style={
    boxrule=0pt, arc=2pt,
    left=6pt, right=6pt, top=2pt, bottom=2pt,
  },
  title={#1},
  toptitle=1pt, bottomtitle=1pt,
}

\begin{document}

%\title{Synthetic Requirements in Experimentation:\\On the Realism of Study Objects}
%\title{Adapt the Object to the Subject:\\Representative Study Material for Empirical SE}
\title{Adapt the Object to the Subject: Representative Study Material for Empirical Software Engineering}

\author{
\IEEEauthorblockN{
Max Unterbusch\IEEEauthorrefmark{1},
Julian Frattini\IEEEauthorrefmark{2},
Michael Mircea\IEEEauthorrefmark{3}, \\
Alexander Korn\IEEEauthorrefmark{1},
Kurt Schneider\IEEEauthorrefmark{3},
Jonathan Ullrich\IEEEauthorrefmark{4}, and
Andreas Vogelsang\IEEEauthorrefmark{1}} \\
\IEEEauthorblockA{\IEEEauthorrefmark{1}\textit{paluno -- The Ruhr Institute for Software Technology, University of Duisburg-Essen}, Germany \\
\{max.unterbusch, alexander.korn, andreas.vogelsang\}@uni-due.de} \vspace{0.15cm}
\IEEEauthorblockA{\IEEEauthorrefmark{2}\textit{Chalmers University of Technology and University of Gothenburg}, Sweden \\
julian.frattini@chalmers.se} \vspace{0.15cm}
\IEEEauthorblockA{\IEEEauthorrefmark{3}\textit{Software Engineering Group, Leibniz University Hannover}, Germany \\
\{michael.mircea, kurt.schneider\}@inf.uni-hannover.de} \vspace{0.15cm}
\IEEEauthorblockA{\IEEEauthorrefmark{4}\textit{Fraunhofer IESE}, Germany \\
jonathan.ullrich@iese.fraunhofer.de}
}

\maketitle
\copyrightnotice

\begin{abstract}
External validity is a central concern in empirical software engineering, yet the debate has largely focused on whether study participants are representative.
How well the study objects they reason about---code, models, or requirements---are representative of each participant's own work is often overlooked.
Practitioners' own artifacts would be ideal, but are often out of reach, so researchers sample from public datasets or construct objects artificially.
Whether this lack of subject–object fit affects participants’ ability to act on the material remains unknown.
We take a first step toward answering this question for requirements.
We propose an approach that uses an LLM to generate context-adapted synthetic requirements just-in-time during an online survey from a short, participant-supplied context description, optionally injecting requirements smells as controlled stimuli.
To validate the approach, we conducted a controlled experiment with 31 industry practitioners in a $2 \times 2$ design that crosses stimulus \emph{generation} (static vs. synthetic) with \emph{quality} (clean vs. smelly).
Each participant rated the perceived representativeness and assessability of 20 requirements.
Using Bayesian data analysis within a statistical causal-inference framework, we find that perceived representativeness has a strong positive effect on assessability.
Synthetic requirements are rated as more representative and---primarily due to this gain---more assessable than static ones sourced from public datasets; smell injection leaves both perceptions unchanged.
Our results support treating a lack of study-object representativeness as a first-class validity threat and position context-adapted synthetic study objects as a practical addition to the empirical methodology toolbox.
\end{abstract}


\section{Introduction}

Empirical studies have become a core instrument of software engineering (SE) research.
As SE is an applied science, it is important that our empirical results are useful to practice, which makes external validity a central concern.
In general, generalizability can be improved by making studies as realistic and representative as possible considering the target population to which one wants to draw conclusions for~\cite{sjobergConductingRealisticExperiments2002}.
Much of this concern has centered on the study participants, where a long-standing debate asks whether students can adequately represent industry practitioners~\cite{salman2015students,falessi2018empirical}.
However, this concern only addresses one dimension of external validity, the interaction of treatment and selection~\cite{wohlinExperimentationSoftwareEngineering2024}.
The extent to which the study objects---including code, commits, architecture diagrams, and requirements---resemble those that a participant would encounter in their own work is often overlooked~\cite{frattini2026towards}.

Using practitioners' own artifacts as study material would have high external validity in the setting dimension, but this is often infeasible:
Such artifacts are often confidential, hard to obtain, and sometimes too sparse in the phenomenon under study to be useful.
A common approach is therefore to sample study objects from public datasets (e.g.,~\cite{scalabrino2018comprehensive, buse2009learning,dalpiaz2019detecting,frattiniApplyingBayesianData2024}) or construct them artificially (e.g., \cite{gopstein2017understanding,basili1996empirical,ajami2019syntax,vogelsangOnTheImpact2025}).
However, in these settings, the material may be of low representativeness and might not match a participant’s domain and expertise.
Currently, we lack evidence on the extent to which the representativeness of study objects affects the assessability by participants.

We take a first step toward answering this question for a particular SE study object, \textit{requirements}.
Requirements are core to many streams of research, including research on requirements quality~\cite{vogelsangOnTheImpact2025,femmerRapidQualityAssurance2017,frattini2024identifying,frattiniApplyingBayesianData2024}.
Unnoticed defects in requirements can become costly problems in downstream SE tasks and risk project success \cite{fernandezNamingPainRequirements2017}.
For this reason, requirements smells have been defined as indicators for potential quality deficiencies~\cite{femmerRapidQualityAssurance2017}.
However, for most smells, we lack empirical evidence for their relevance~\cite{frattini2024identifying, montgomeryEmpiricalResearchRequirements2022a}, although first studies have begun to investigate their effects~\cite{gentiliPractitionersPerceptionsRequirements2025, laprovaAnticipatingBugsTicketlevel2026,frattiniApplyingBayesianData2024}.
We believe part of the reason is that requirements quality is inherently difficult to study.
Whether a requirement smell is genuinely harmful depends on context factors, such as domain, the process model, or the availability of stakeholders who could resolve it~\cite{frattini2024identifying}.
Experiments in requirements quality research are only meaningful if participants can relate the presented requirements to a context they know; confronting them with unfamiliar material would undermine the judgments the study relies on.

To address this problem, we propose and evaluate an approach that generates \textit{context-adaptive synthetic requirements} just-in-time and uses them as study objects.
At the start of a study, each participant supplies a short description of their working context, which is used to condition a large language model (LLM) that generates requirements tailored to that participant on the fly.
To evaluate whether the approach is capable of adding controlled stimuli, an additional LLM-based mutation step injects requirements smells into the generated study objects, yielding both clean and smelly requirements.

We conducted a controlled experiment with 31 industry practitioners, following a $2\times2$ within-subjects design that crosses the stimulus \textit{generation} (synthetic vs.\ static) with the requirement \textit{quality} (clean vs.\ smelly).
The \emph{static} condition serves as the control level: requirements sampled from projects in public datasets without adaptation to the participant.
Each participant rates requirements from all four combinations of conditions on two questions, capturing the dimension of perceived \textit{representativeness} and \textit{assessability}.
Both constructs are first-person judgments that only the practitioners themselves can make in their context.

Analyzing the ratings with Bayesian data analysis~\cite{mcelreath2018statistical} within a framework for statistical causal inference~\cite{siebert2023applications}, we find that perceived representativeness has a strong positive effect on participants' assessability, i.e., their confidence in acting on the requirements.
Since representativeness could not be experimentally manipulated, we adjusted for known confounders in our causal model and show in a sensitivity analysis that the observed association could only be spurious if an unknown confounder were implausibly strong.
Synthetic requirements, which we experimentally controlled, are rated as more representative and, primarily for this reason, more assessable than static ones sourced from public datasets.
The LLM-based smell mutation leaves both perceptions unchanged.

In summary, this paper makes the following contributions:
\begin{itemize}[leftmargin=*, topsep=2pt]
    \item \textbf{Empirical evidence} that study-object representativeness has a strong positive effect on practitioners' perceived ability to act on the material.
    Demonstrated here for requirements, this supports treating a lack of object representativeness as a first-class validity threat in human-subject experiments and motivates studying it for other types of study objects.
    \item \textbf{An approach} for generating \textit{context-adapted study objects} just-in-time from a short, participant-supplied context description, here instantiated for requirements with an optional LLM-based mutation step that injects requirements smells. In a controlled experiment with 31 practitioners, synthetic requirements were perceived as more representative and, primarily through this gain, more assessable than static ones sampled from public datasets, while smell injection left both perceptions unchanged. The approach is in principle transferable to other types of SE study objects.
    \item \textbf{Methodological guidance} on when the approach is and is not appropriate, and how to apply it.
\end{itemize}

\section{Background \& Related Works}
\label{sec:background}

\subsection{Validity of Experiments}
\label{subsec:validity}

The validity of an experiment is a fundamental concern in empirical research.
Following Wohlin et al.~\cite{wohlinExperimentationSoftwareEngineering2024}, adequate validity is achieved when the experimental results hold for the target population to which the experimenters intend to generalize.
Four core types of validity are typically considered: Conclusion, construct, internal, and external validity.
These types of validity usually cannot be maximized simultaneously within a single experiment and must therefore be prioritized with respect to the experiment's purpose~\cite{wohlinExperimentationSoftwareEngineering2024}.

Specifically important to this work are internal validity, which concerns the ability to draw causal conclusions between the variables of interest, and external validity, which concerns the ability to generalize the results beyond the study setting~\cite{wohlinExperimentationSoftwareEngineering2024}.
While experiments focus on establishing causal evidence with high internal validity, external validity must not be neglected.
As Sj{\o}berg et al.~\cite{sjobergConductingRealisticExperiments2002} argue, an experiment that lacks external validity is of no practical use, since its conclusions hold only within the experimental setting.
Hence, they identify realism as a particular concern for generalizability to industrial practice and recommend that experimental tasks should be as representative of industrial practice as possible.

It is important to be aware that external validity is a multi-dimensional concept.
Wohlin et al. distinguish three types, defined by the interaction of the treatment with selection, history, and setting~\cite{wohlinExperimentationSoftwareEngineering2024}.
The \textit{selection} dimension---most strongly associated with external validity for many SE researchers---refers to validity threats that occur because the population of study subjects is not representative of the population to which the experimenters intend to generalize.
%The extensive debate about how well students represent SE practitioners revolves around this dimension~\cite{salman2015students,falessi2018empirical}.
However, there are also other dimensions of external validity.
The \textit{history} dimension refers to threats that occur because the experiment is preceded by special events that might impact the results.
Most important for this work is the \textit{setting} dimension, the interaction of setting and treatment, which Wohlin et al. define as ``the effect of not having the experimental setting or material representative of, for example, industrial practice''~\cite[p. 138]{wohlinExperimentationSoftwareEngineering2024}.

Regarding the first-mentioned dimension of \textit{selection}, a widely discussed validity concern in SE research is the choice of study participants~\cite{salman2015students,falessi2018empirical}.
It is broadly acknowledged that experiments with professional participants are desirable, and there is evidence that practitioners perceive studies conducted with industry professionals as more relevant~\cite{franchHowPractitionersPerceive2022}.
In practice, however, this is often difficult to realize as the availability of professionals is limited.
The choice of study subjects thus represents a deliberate tradeoff: employing practitioners increases external validity but, compared to the more homogeneous population of students, comes at the cost of control~\cite{sjobergConductingRealisticExperiments2002}.

A less often discussed threat to external validity stems from the \emph{study objects}~\cite{frattini2026towards}, which are part of the \textit{setting} dimension.
Context-fit, i.e., how well the study objects fit a study participant's context, may moderate the effect of a stimulus on an outcome.
To illustrate the importance of using representative study objects, consider the following example from requirements quality research:
In an experiment where students reviewed a static set of requirements unfamiliar to them, the use of passive voice (the stimulus, which was supposedly associated with ambiguity/poor quality) greatly affected a subsequent modeling task (the outcome)~\cite{femmer2014impact}.
However, in a different study conducted at a case company, passive voice turned out to be practically harmless~\cite{krisch2015myth}.
An important difference between the studies is that in the latter, the context-fit between experiment subjects and objects was much greater as practitioners are more familiar with their requirements.
Hence, passive voice was rarely judged problematic in the actual context because the practitioners' domain knowledge and working context compensated the supposed defect~\cite{krisch2015myth}.

In the example above, the latter study has more external validity with respect to the setting dimension than the former because it had access to the \textit{real} requirements from their subject's working context.
In research practice, however, real artifacts are often out of reach---most commonly due to confidentiality, but also when the phenomenon under study occurs too rarely for any single practitioner to supply enough examples, or when the material does not yet exist, as when researchers evaluate a newly proposed SE method.

While prior work suggests that domain familiarity aids RE tasks~\cite{hadarRoleDomainKnowledge2014}, we lack evidence on whether and to what extent the representativeness of study objects affects the assessability by participants.
In this confirmatory research, we test and quantify this effect in a controlled experiment and derive implications for the design of experiments.

\subsection{Requirements Generation with LLMs}

LLM-assisted requirements elicitation and documentation has become an active area of requirements engineering (RE) research~\cite{zadenooriLargeLanguageModels2025}. 
Several studies have evaluated the quality of LLM-generated requirements against human baselines.
Krishna et al.~\cite{krishnaUsingLLMsSoftware2024a} find that GPT-4-2023-09-25 and CodeLlama-34b produce software requirements specifications comparable in completeness and consistency to those of entry-level engineers.
They also find that the generated requirements largely avoid ambiguities.
Quattrocchi et al.~\cite{quattrocchiCanLLMsGenerate2026} investigate user story generation across multiple LLMs and report similar coverage and stylistic quality to human authors, though with less creativity and diversity regardless of model size.
Beyond generation from scratch, Mircea et al.~\cite{mirceaSupportingStakeholderRequirements2026} show that LLM revisions of stakeholder-authored requirements improve stakeholder alignment, readability, and unambiguity.

A smaller body of work uses LLM-generated requirements for research purposes rather than practitioner support.
Fantechi et al.~\cite{fantechiInconsistencyDetectionNatural2023} generated conflicting requirements from a seed set using ChatGPT and subsequently asked the model to detect the resulting inconsistencies.
El-Hajjami and Salinesi~\cite{el-hajjamiSynthlineProductLine2025} evaluate synthetic requirements as training data for ML classifiers, finding that despite lower diversity, they are valuable for classification tasks.
Femmer et al.~\cite{femmerDescriptionComparativeAnalysis2025a} investigated linguistic properties of requirements generated by gpt-4.1-2025-04-14 under varying amounts of contextual input, finding that richer context improves output quality, yet synthetic requirements remain syntactically simpler than those in public datasets.

Our work differs from both strands.
The practitioner-facing studies above compare LLMs and humans as competing \textit{producers} of requirements for the same project, while the research-oriented studies feed synthetic requirements into ML models or other automated tasks.
Neither evaluates synthetic requirements as \textit{study objects} for human participants in empirical studies.
The use of LLMs to generate experimental material has also been proposed in other fields, notably psychology~\cite{behrend2025participant}.
To our knowledge, however, this approach has neither been proposed nor examined for SE study objects.

\section{Study Design}

We report our study design following the guideline for reporting controlled experiments by Jedlitschka et al.~\cite{Jedlitschka2008}.

\subsection{Study Objectives, Research Questions \& Overview}

The first objective of this paper is to investigate whether the representativeness of a requirement is itself a determinant of practitioners' perceived ability to act soundly on the material.
We consider a requirement \textit{representative} when a practitioner could plausibly encounter it in their own work, and \textit{assessable} when they feel confident performing their typical tasks on it.
Establishing such a link is the question on which the importance of all other representativeness-related design choices in human-subject studies depends.
We raise the following research question (RQ):

\begin{enumerate}[label=\textbf{RQ\arabic*.}, leftmargin=*, ref=RQ\arabic*]
    \item To what extent does perceived \textit{representativeness} of a requirement affect its perceived \textit{assessability}?
\end{enumerate}

The second objective of this paper is to investigate the general suitability of LLM-generated requirements to be used as context-adapted study objects in empirical research.
We investigate whether practitioners, as we hypothesize, rate the synthetic requirements as more representative (important for external validity in the setting dimension, see Section~\ref{subsec:validity}) and as more assessable (important for the soundness of experiments) than the static baseline.
We thus ask:
\begin{enumerate}[label=\textbf{RQ\arabic*.}, leftmargin=*, ref=RQ\arabic*, start = 2]
    \item To what extent do practitioners perceive synthetic requirements as \textit{representative} compared to static ones?
    \item To what extent do practitioners perceive synthetic requirements as \textit{assessable} compared to static ones?
\end{enumerate}

The third objective is to investigate whether synthetically generated requirements can be further mutated to inject requirements smells while preserving their perceived representativeness and assessability.
This invariance is a precondition for using the pipeline as a source of study stimuli: most empirical studies must manipulate their study objects to administer a treatment.
If the smell injection itself altered perceived representativeness or assessability, the treatment would be confounded with these properties, rendering the approach useless for this purpose.
Hence, we also ask:
\begin{enumerate}[label=\textbf{RQ\arabic*.}, leftmargin=*, ref=RQ\arabic*, start=4]
    \item Does LLM-based smell mutation affect the perceived \textit{representativeness} of synthetic requirements?
    \item Does LLM-based smell mutation affect the perceived \textit{assessability} of synthetic requirements?
\end{enumerate}

\begin{figure}
    \centering
    \includegraphics[width=0.3\linewidth]{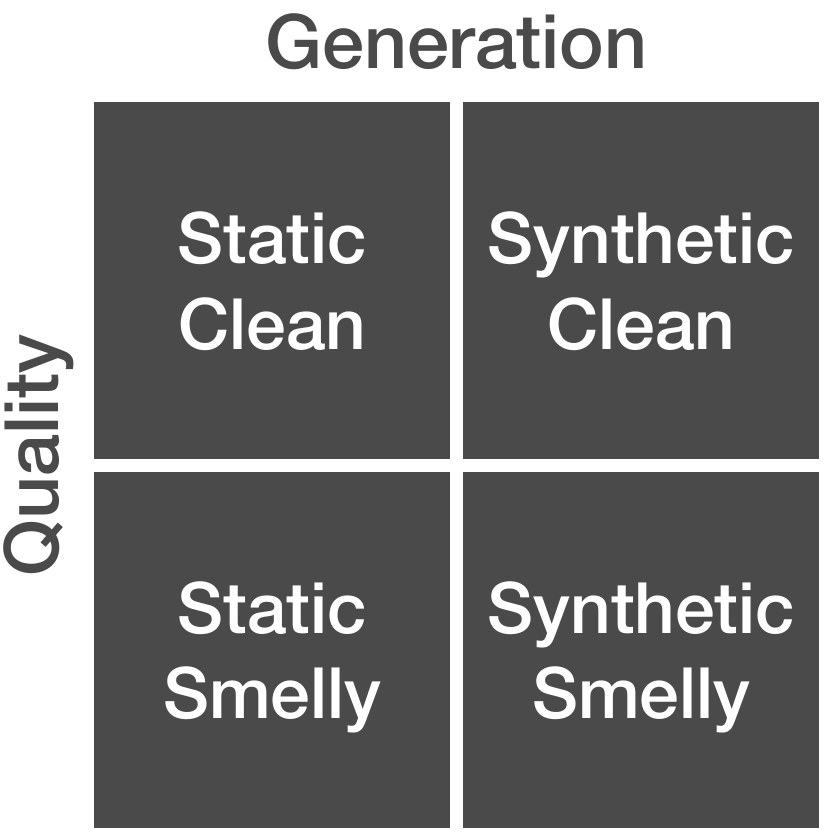}
    \caption{Conditions of the $2 \times 2$ within-subjects design.}
    \label{fig:conditions}
\end{figure}

To address these RQs, we conduct a controlled experiment with 31 industry practitioners.
The study follows a $2\times2$ within-subjects design that crosses the factor \textit{generation} (synthetic vs.\ static) with the factor \textit{quality} (clean vs.\ smelly), yielding four conditions under which each participant evaluates five distinct requirements (see Figure~\ref{fig:conditions}).
Both levels of the \textit{generation} factor represent a subset of the requirements from a single coherent project, shown to the participant in a randomized order, yet only the synthetic condition is context-adapted.
In the \textit{synthetic} condition, the project is generated on the fly by an LLM from the participant's self-reported context (application domain and system description).
The \textit{static} condition serves as control level, where the project is randomly selected from a public requirements dataset with no adaptation to the participant, as often done in RE research (e.g.,~\cite{frattiniApplyingBayesianData2024,dalpiaz2019detecting}).
The \textit{static-smelly} condition contains naturally occurring smells, while the \textit{synthetic-smelly} requirements have been injected by an LLM.

We specifically chose a within-subjects design because between-subjects variance among industry practitioners is expected to be large relative to our feasible sample size.
Pairing the four conditions within each participant therefore yields substantially more statistical power than a between-subjects design at the same $N$.
Simultaneously, the within-subjects design helps avoid the tendency of participants to respond neutrally to Likert-scale items due to the lack of comparison in a between-subjects design.

\subsection{Study Objects}

\subsubsection{Static Requirements}
To benchmark context-adapted against static study objects, we draw the static requirements from two publicly available datasets, each covering one popular requirements format:
For classical specification-style requirements, we use the PROMISE+ dataset~\cite{carvalhoPromise2024} containing 88 projects and for user stories we use the dataset provided by Dalpiaz~\cite{dalpiazUserStories2018}, containing 22 projects.

Since our design also compares clean and smelly study objects, we cannot sample arbitrarily but have to ensure that the static condition contains genuine, naturally occurring smells.
We focus on five of the nine smell types used by Vogelsang et al.~\cite{vogelsangOnTheImpact2025}, originally drawn from the catalog of Frattini et al.~\cite{frattiniLiveExtensibleOntology2022}: \textit{subjective language}, \textit{weak verbs}, \textit{optional parts}, \textit{negative statements}, and \textit{passive voice}.
We excluded three smells that are inherently relational (e.g., contradiction) because we require every smell to be self-contained in a single requirement.
We later also excluded one smell that occurred only rarely in the two datasets (vague pronoun).

We then systematically mined both datasets for projects that contain examples of \textit{each} of the smells as well as a sufficient number of clean requirements.
For all of the 5,357 requirements, we individually prompted an efficient OpenAI model (gpt-5.4-mini-2026-03-17 with temperature~0) to rate the presence of each smell on an ordinal scale from 0 (confident not present) to 4 (confident present).
The full prompt is provided in our online appendix\footnote{\url{http://doi.org/10.6084/m9.figshare.32397573}}.
A project qualified for our study if it contained (i)~at least one requirement with confidence 4 for each of the five smells and (ii)~at least five requirements with confidence 0 across all smell categories.
Whenever there were multiple candidates per class (e.g., requirement with subjective language smell), we selected the requirement at random.
This yielded five qualifying user-story and specification-style projects each.

To validate the LLM-based pre-filtering, two authors independently labeled each candidate (smell genuinely present~/ requirement genuinely clean).
Both raters agreed on 94 of 100 cases, corresponding to Cohen's $\kappa = 0.67$ (substantial) and Gwet's $\text{AC1} = 0.93$ (very strong).
In 13 cases at least one annotator rejected the candidate.
For these cases, we drew a replacement requirement from the same project on which both raters agreed. 
In three cases no suitable replacement within the same project could be found for the originally targeted smell; we substituted a requirement bearing a different smell and balanced these substitutions across \textit{subjective language}, \textit{negative statement}, and \textit{optional part} (one each).

\subsubsection{Synthetic Requirements}
\label{subsubsec:synthetic_requirements}
For the synthetic requirements condition, we generate 20 requirements with OpenAI's flagship model at the time of writing (gpt-5.5-2026-04-23) conditioned on the participant's context (application domain and system description) just-in-time while our participants are filling the survey.
We used a temperature of 0.7 to balance output determinism and creativity in the generated requirements.
The prompt is shown in Figure~\ref{fig:generation-prompt}.

\begin{figure}
\begin{promptbox}[Prompt: Requirements Generation]
\footnotesize\ttfamily
\promptlabel{Context} You are an expert requirements engineer in
\promptvar{domain}. You value high-quality requirements.

\smallskip
\promptlabel{Task} Generate 20 requirements for the following system:
\promptvar{systemDescription}. The set of requirements does not need
to be complete. Write the requirements in the following style:
\promptvar{requirementsFormat}.

\smallskip
Your output MUST be a valid JSON in the format
\{REQ1: requirement, \ldots, REQ20: requirement\}.
\end{promptbox}
\caption{\small Prompt template used to generate context-adapted synthetic
requirements. Runtime placeholders are shown in
\textcolor{promptVar}{\textbf{[\dots]}}.}
\label{fig:generation-prompt}
\end{figure}

We feed 15 of the 20 generated requirements (positions 2--4, 6--8, 10--12, 14--16, 18--20) into a second LLM call that mutates them to introduce smells for the synthetic-smelly cell.
We prompt the same model to produce exactly five mutations, one per smell type, and the model is free to choose \textit{which} of the 15 candidates to mutate, so that no requirement is forced to carry a smell that would be unnatural for it.
The mutation prompt is shown in Figure~\ref{fig:mutation-prompt}.
The remaining five requirements (positions 1, 5, 9, 13, 17, i.e., every fourth requirement) are passed through unchanged and form the synthetic-clean cell.
Generating initially only 20 requirements is a deliberate trade-off to keep total waiting time below roughly 15 seconds.
Consequently, the synthetic requirements that the participant eventually sees are also a non-contiguous excerpt of their underlying project, matching the static condition in this regard.

\begin{figure}
\begin{promptbox}[Prompt: Smell Mutation]
\footnotesize\ttfamily
\promptlabel{Context} You are an expert requirements engineer in
\promptvar{domain} and you will be given 15 requirements for the
following system: \promptvar{systemDescription}.

\smallskip
\promptlabel{Task} Modify some of the requirements by adding
requirements smells. For each of the following five smells, pick
one requirement out of the 15 where you can apply it: \textit{subjective
language, optional parts, weak verbs, passive voice, and negative
phrases (each smell is provided with a definition and two
examples).}

\smallskip
\promptlabel{Requirements} \promptvar{REQ2-REQ4}, ..., \promptvar{REQ18-REQ20}

\smallskip
Your output MUST be a valid json, \textit{keyed by REQSUBJ, REQOPT, REQWEAK, REQPASS, REQNEG.}
\end{promptbox}
\caption{\small Prompt template used to mutate five of the 15 supplied requirements to contain smells. Runtime placeholders are shown in
\textcolor{promptVar}{\textbf{[\dots]}}; content abbreviated for
space is shown in italics.}
\label{fig:mutation-prompt}
\end{figure}

\subsection{Study Execution}

\subsubsection{Instrumentation}
The study is implemented as an online survey in LimeSurvey, extended with the LLM Survey Plugin that allows us to trigger LLM calls just-in-time between pages based on participant input\footnote{\url{https://git.se.uni-hannover.de/llm-survey-tool/llmsurvey}. Version~2.3 of the plugin is also available in our online material\textsuperscript{1}}.
The survey has four pages:

\textbf{Page~1 -- Context elicitation.}
Participants provide the three inputs that drive the generation step: the \textit{application domain} (free text, e.g., ``consumer mobile app''), a \textit{system description} of one or two sentences (free text, e.g., ``a language-learning app with gamification elements for motivation''), and the \textit{requirement format} most representative of their work (one of \textit{classical specification-style} or \textit{user story}).
In addition, participants select the activities they typically perform on the basis of requirements (e.g., \textit{writing requirements}, \textit{writing code}, \textit{deriving
test cases}).
This last question primes participants for the assessability item on Page~3 and lets us interpret their answers relative to their familiar working tasks.
While the participant transitions from Page~1 to Page~2, two things happen in the background.
First, the LLM generates 20 requirements (see Section~\ref{subsubsec:synthetic_requirements}), which takes approximately 10 seconds, during which the participants see a loading animation.
Second, we randomly draw one of the five curated projects of static requirements that match the chosen format.

\textbf{Page~2 -- Task description.}
Page~2 introduces the rating task and shows the two rating items.
On the transition from Page~2 to Page~3, we make the LLM call to mutate the previously generated requirements to contain smells (see Section~\ref{subsubsec:synthetic_requirements}), which takes about five seconds.
We chose to have this intermediate screen between the generation and mutation steps in order to avoid a single, yet longer waiting screen.

\textbf{Page~3 -- Rating task.}
Each participant evaluates 20 requirements, five per each of the four cells (see Figure~\ref{fig:conditions}), presented in randomized order and fully blinded.
Participants were told neither that some requirements were synthetically generated nor that some were deliberately mutated to carry a smell.
For each stimulus, participants respond to two 5-point Likert items on a scale from \textit{strongly disagree} (1) to \textit{strongly agree} (5).
The first item, \textit{``I could imagine encountering this requirement in my own work,''} operationalizes \textit{representativeness}.
The second item, \textit{``I feel confident that I could perform my typical tasks based on this requirement,''} operationalizes \textit{assessability} as the self-reported ability to act soundly on the requirement.

\textbf{Page~4 -- Closing.}
Finally, we elicit demographic information to characterize the sample: years of professional experience in software or systems engineering (integer estimate), frequency of working with requirements artifacts (\textit{daily}, \textit{weekly}, \textit{monthly}, \textit{rarely}, \textit{never}), and a self-positioning between \textit{exclusively academic} (1) and \textit{exclusively industry} (5) on a five-point scale.
An optional free-text field invites participants to share general comments.

Before launch, we piloted the complete survey with three non-author participants.
The pilot confirmed that the instrument functioned as intended and, in particular, that the two rating items were interpreted consistently with our intention.

\subsubsection{Participants \& Data Collection}

We aimed to recruit 30 industry practitioners with a software or systems engineering background via personal invitations.
We consider this sampling strategy appropriate and refrained from openly calling for participants in order to avoid academic peers diluting our sample.
The intended target group is practitioners who routinely engage with written requirements artifacts; we therefore filter out respondents who report \textit{never} working with requirements, as well as those that report having an exclusively academic background.
The survey was open from May to June 2026 and participation was entirely voluntary.

\subsubsection{Requirements Analysis}

Previous works showed that LLM-generated requirements may have lower syntactical complexity than real requirements~\cite{femmerDescriptionComparativeAnalysis2025a}.
Such under-complexity would limit the realism of the study objects~\cite{frattini2026towards,sjobergConductingRealisticExperiments2002}.
To investigate whether our synthetic requirements exhibit the same under-complexity, after data collection, we analyze metrics of the previous study: parse-tree-depth, defined as the longest path from root to leaf in the dependency tree of a sentence~\cite{brunato2018sentence} and the sentence length measured in the number of lexical words of the requirements.
Moreover, to validate that the smell mutation step produced the intended smell types, two authors independently re-inspect each \textit{synthetic-smelly} requirement.

\subsection{Data Analysis}

We utilize Bayesian data analysis (BDA)~\cite{mcelreath2018statistical} within a framework for statistical causal inference (SCI)~\cite{siebert2023applications} to answer our RQs based on the collected data.
We chose BDA over the more common frequentist methods for their fine-grained insights~\cite{furia2019bayesian} and their ability to properly model data that is not normally distributed~\cite{gren2021possible}.
Since our outcome variables are on an ordinal scale and, hence, not normally distributed, BDA is more eligible.
The SCI framework allows differentiating causal relationships from spurious associations~\cite{pearl2009causal}.
We adhere to the framework by Siebert~\cite{siebert2023applications} comprising a \emph{modeling}, \emph{identification}, and \emph{estimation} step.

\subsubsection{Modeling}

\begin{figure}
    \centering
    \includegraphics[width=\linewidth]{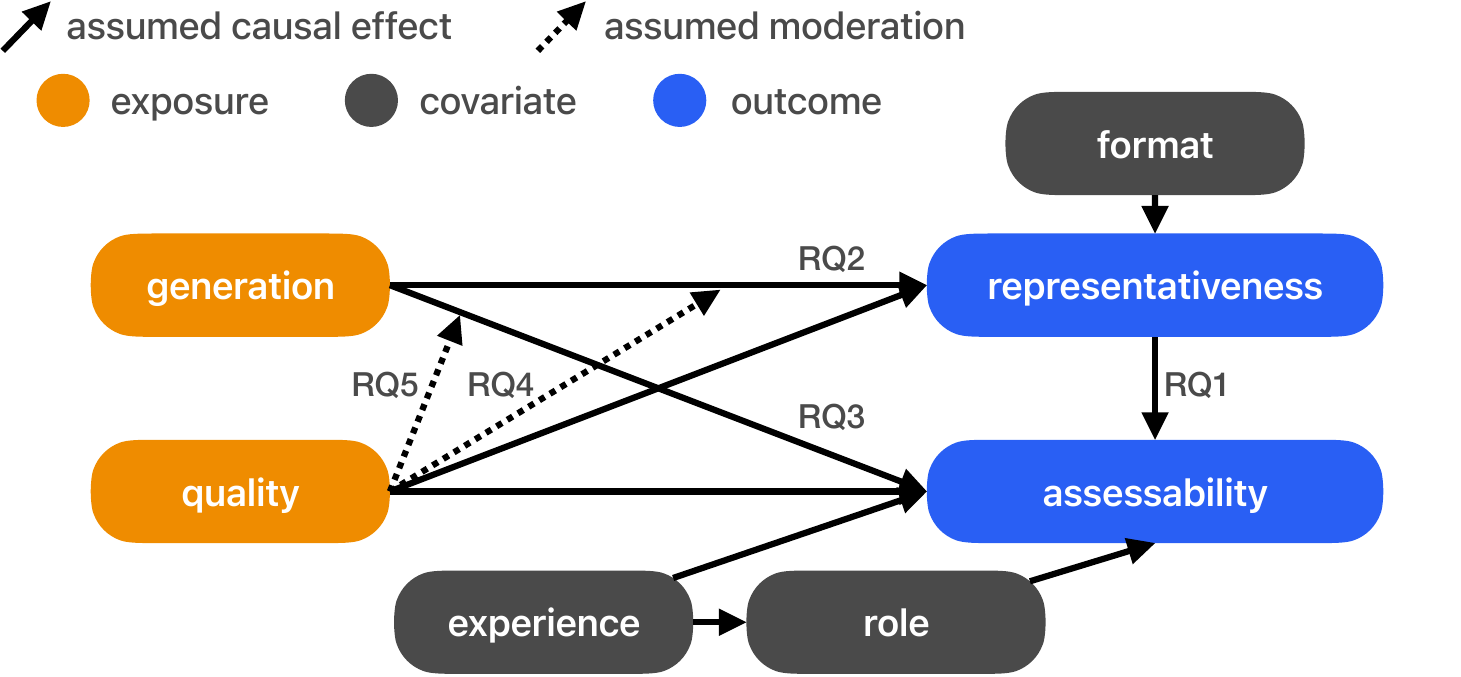}
    \caption{DAG encoding our causal assumptions}
    \label{fig:dag}
\end{figure}

First, we explicitly specified our causal assumptions in the form of a directed acyclic graph (DAG)~\cite{glymour2008causal}.
In the resulting DAG, visualized in \Cref{fig:dag}, every node represents a variable and every directed edge an assumed causal relationship.
For example, we assumed that the \texttt{representativeness} of a requirement affects its \texttt{assessability} (RQ1).
\Cref{tab:variables} contains a complete list of our variables.
In our online material,\textsuperscript{1} we justified each included edge between two variables, which we omitted from this manuscript for brevity.

\begin{table*}[hbt]
    \centering
    \caption{Relevant variables involved in the study and their role in the
    \textit{representativeness} (Rep.) and \textit{assessability} (Ass.) models\\
    ($\circ$ = outcome, $\bullet$ = exposure, \textcolor{gray}{$\bullet$} = covariate).
    Both models additionally include the interaction
    \texttt{generation}\,$\times$\,\texttt{quality}.}
    \label{tab:variables}
    \begin{tabularx}{\textwidth}{lXllcc}
        \toprule
        \textbf{Variable} & \textbf{Description} & \textbf{Type} & \textbf{Scale} & \textbf{Rep.} & \textbf{Ass.} \\
        \midrule
        \texttt{generation} & The method with which the requirement was generated & categorical & \{static = 0, synthetic = 1\} & $\bullet$ & $\bullet$ \\
        \texttt{quality}  & Whether or not the requirement contains a smell & categorical & \{clean = 0, smelly = 1\} & $\bullet$ & $\bullet$ \\
        \midrule
        \texttt{format}  & The format in which a requirement is specified & categorical & \{specification, user story\} & \textcolor{gray}{$\bullet$} & \textcolor{gray}{$\bullet$} \\
        \texttt{experience}  & The number of years of experience of the participant & continuous & $\mathbb{N}$ & & \textcolor{gray}{$\bullet$} \\
        \texttt{role}  & The activities that a participant performs on requirements & categorical & \{write, design, code, ...\} & & \textcolor{gray}{$\bullet$} \\
        \midrule
        \texttt{representativeness}  & How representative a participant perceives a requirement to be & ordinal & \{1, 2, 3, 4, 5\} & $\circ$ & $\bullet$ \\
        \texttt{assessability} & How assessable a participant perceives a requirement to be & ordinal & \{1, 2, 3, 4, 5\} & & $\circ$ \\
        \bottomrule
    \end{tabularx}
\end{table*}

\subsubsection{Identification}

From our causal assumptions modeled in a DAG, we identified the adjustment set, i.e., the set of variables necessary to statistically control to avoid bias via confounding~\cite{mcelreath2018statistical}.
Since the data were collected via a controlled experiment and the two exposure variables were fully controlled, confounders were eliminated by design for RQ2-RQ5~\cite{pearl2009causal}.
Still, we statistically adjusted for all covariates affecting any outcome variable to improve the precision of the subsequent estimation~\cite{cinelli2024crash}.

\subsubsection{Estimation}

Finally, we estimated the strength of the assumed causal relationships as specified in \Cref{fig:dag}.
For this, we trained two Bayesian regression models---one per outcome variable---following established procedures for BDA~\cite{mcelreath2018statistical}.

For every regression model, we choose the outcome distribution that is consistent with what we know about the outcome variable while adding no further assumptions~\cite{jaynes2003probability}.
In our case, both outcome variables are measured on an ordinal scale with five levels.
This implies that (1) any valid value must be one of the five levels, (2) the five levels have an inherent order to them, but (3) the distances between adjacent levels are unknown and not necessarily equal (e.g., the distance from level 1 to 2 need not equal that from 2 to 3).
We therefore use a cumulative model~\cite{burkner2019ordinal} (\Cref{eq:likelihood}) to represent the outcome variable.
The cumulative model assumes that each response results from an underlying, latent variable, which can be understood as the continuous \emph{attitude} of a respondent towards a statement.
In the absence of contrary evidence, we assume that this latent attitude is normally distributed, i.e., it follows the maximum-entropy distribution for a real-valued quantity with a given mean and variance.
The location $\phi_i$ of this distribution is described by a linear predictor (\Cref{eq:model1,eq:model2,eq:model3}), which specifies how the predictors shift a participant's latent attitude toward \emph{representativeness} (and, equivalently, toward \emph{assessability}).
The thresholds $\kappa_k$ are the values of the latent attitude at which a participant switches from one response level (e.g., ``neutral'') to the next (e.g., ``agree'').
In summary, a respondent selects option $k$ or lower if their latent attitude, which is normally distributed around $\phi_i$, falls below the threshold $\kappa_k$, and an option above $k$ otherwise.
Formally, $\Pr(\texttt{rep}_i \le k) = \Phi(\kappa_k - \phi_i)$, where $\Phi$ is the cumulative distribution function of the standard normal distribution (a so-called probit link).

\begin{align}
    \texttt{rep}_i \sim &~\text{Cumulative}(\phi_i, \kappa) \label{eq:likelihood}\\
    \phi_i = &~\alpha_{ID[i]} + \beta_g \cdot \texttt{gen}_i + \beta_q \cdot \texttt{quality}_i  \label{eq:model1}\\
      &+ \beta_{g \times q} \cdot \texttt{gen}_i \cdot \texttt{quality}_i  \label{eq:model2}\\
      &+\beta_f \cdot \texttt{format}_i \label{eq:model3}\\
    \alpha_j \sim &~\mathcal{N}(0, \sigma) \quad \text{for } j = 1, \dots, J \label{eq:prior:cluster}\\
    \sigma \sim &~\text{Weibull}(2, 1) \label{eq:prior:cluster:sigma}\\
    \beta_g, \beta_q \sim &~\mathcal{N}(0, 0.5) \label{eq:prior:coef:main}\\
    \beta_{g \times q}, \beta_f \sim &~\mathcal{N}(0, 0.2) \\
    \kappa_k \sim &~\mathcal{N}(0, 1) \quad \text{for } k = 1, \dots, K-1, \quad \\
     &~\kappa_1 < \dots < \kappa_{K-1}
\end{align}

The linear predictor contains a participant-specific varying intercept $\alpha_{ID[i]}$ (\Cref{eq:prior:cluster,eq:prior:cluster:sigma}) and a coefficient for each predictor (e.g., $\beta_q$ for \texttt{quality}).
Because every participant provided several ratings, the varying intercept accounts for the resulting non-independence and captures systematic, unmeasured differences between participants, such as a general tendency to agree.
The varying intercepts regularize these estimates, which typically improves estimation compared to models that ignore the clustering~\cite{ernst2018bayesian}.
The coefficients $\beta$ are the quantities of interest for the analysis.
They are expressed on the latent (probit) scale, i.e., in standard deviations of the latent variable.

Given the lack of strong evidence about these effect strengths, we select uninformative prior probability distributions to initialize all coefficients~\cite{mcelreath2018statistical}.
For example, we initialize $\beta_g$ with a normal distribution centered around 0 and with a standard deviation of 0.5 (\Cref{eq:prior:coef:main}). 
This distribution puts equal probability mass above and below 0, representing our prior belief that the effect of the \texttt{generation} method could be either negative or positive.
We performed prior predictive checks by sampling from these priors directly without updating them to confirm the eligibility of the priors~\cite{wesner2021choosing}.

Once confirmed, we trained the model using Markov Chain Monte Carlo sampling~\cite{brooks2011handbook}.
This process updates the coefficients in light of the collected data to better reflect the observed effects of the predictors on the outcome variables, producing posterior probability distributions.
After training, we checked diagnostics~\cite{mcelreath2018statistical} and performed posterior predictive checks~\cite{wesner2021choosing} to confirm that the regression models trained well. 
At this point, the coefficients represent the likely strength of the effects that the predictors have on the outcomes, under the assumption specified in \Cref{fig:dag} and our prior beliefs.

This allows answering our RQs from the posterior probability distributions.
For all RQs, we check whether the 95\% credible interval (CI) of the coefficient associated with the factor in question contains 0.
If not, then it implies a posterior probability of at least 97.5\%\footnote{The equal-tailed 95\% CI leaves 2.5\% of posterior mass beyond each bound. If the lower bound is above 0, the 95\% inside the interval and the 2.5\% above it are both positive, so at least 97.5\% of posterior mass is $\geq0$.} that the factor affects the respective outcome variable either positively (CI $>$ 0) or negatively (CI $<$ 0).
%For easier interpretation, we label such effects \emph{credibly positive/negative}.
%This decision rule plays a role similar to a significance test in frequentist analyses, but it is not equivalent:  
%it is a statement about the posterior probability of the effect given the data and priors, not about the probability of the data under a null hypothesis~\cite{mcelreath2018statistical}.
%Hence, we encourage readers to consider the full posterior probability distributions, which also convey the magnitude and uncertainty of each effect.

RQ1 requires inferring about the effect of perceived \texttt{representativeness} on perceived \texttt{assessability}.
This effect is captured by the path coefficient of \texttt{representativeness} in the \texttt{assessability} model, $\beta_{represent}$, with \texttt{generation} adjusted as a confounder so that the coefficient admits a causal interpretation (see Figure~\ref{fig:dag}).
RQ2 and RQ3 require inferring about the effect of the \texttt{generation} approach on the \texttt{representativeness} (RQ2) and \texttt{assessability} (RQ3) of requirements.
These effects are captured by the path coefficient of the \texttt{generation} approach $\beta_{g}$ in the respective regression model.
%To contextualize the answers, we also visualize the marginal effect of the exposure variable on the outcomes.
%The marginal effect is computed by varying only one variable (here: \texttt{generation}) while keeping all other predictors constant at representative levels, therefore isolating the effect of the selected variable on the outcomes.
RQ4 and RQ5 require inferring about how requirements \texttt{quality} moderates the effect of the \texttt{generation} approach on the \texttt{representativeness} (RQ4) and \texttt{assessability} (RQ5) of requirements.
Similarly, we investigate the 95\% CI of the path coefficient of the interaction effect $\beta_{g \times q}$.

\subsubsection{Validity Assessment}

The internal validity of the result to RQ1 may be threatened, as this is the only RQ where the exposure (representativeness) was not experimentally manipulated.
Hence, the effect of representativeness on assessability may be biased by unobserved confounders~\cite{cinelli2020making,furia2026mitigating}.
The participant-specific intercept $\alpha_{ID[i]}$ already absorbs any confounder that is constant within a participant, such as a general tendency to rate leniently.
A remaining confounder would therefore have to vary across the requirements that a participant rates, such as the clarity of a requirement.

To assess the reliability of the results to RQ1, we perform a sensitivity analysis~\cite{lin1998assessing}.
We consider a hypothetical confounder $U$, standardized to unit variance, that is characterized by two parameters:
$\gamma$, the effect of a one-standard-deviation increase in $U$ on the latent assessability, and $\delta$, the difference in the mean of $U$ between requirements rated lowest and highest in representativeness.
Omitting such a confounder from the model biases the estimated effect of representativeness on assessability by approximately $\gamma \cdot \delta$~\cite{lin1998assessing}.
We express the effect of representativeness as $\Delta$, the difference in latent assessability between the lowest and the highest level of representativeness, which combines all polynomial contrasts~\cite{burkner2019ordinal}.
For a grid of values of $\delta$ and $\gamma$, we subtract the bias $\gamma \cdot \delta$ from every posterior draw of $\Delta$ and determine the combinations at which the lower bound of the 95\% credible interval reaches zero.
These combinations describe how strong a confounder would have to be to render the observed effect compatible with zero.
Vice versa, this would be the necessary strength of the confounder in order to suggest the observed effect strength if it were indeed not there.

Since absolute values of $\delta$ and $\gamma$ are hard to judge, we compare them to two empirical anchors from our own data~\cite{cinelli2020making}.
For $\gamma$, we use the standard deviation $\sigma$ of the participant-specific intercepts $\alpha_{ID[i]}$ in the assessability model.
A confounder with $\gamma = \sigma$ would affect assessability as strongly as the entire variation between participants in their baseline assessability ratings.
For $\delta$, we use $\delta_{syn}$, the imbalance of the experimentally manipulated factor \emph{generation} between requirements rated lowest and highest in representativeness, standardized in the same way as $U$.
A confounder with $\delta = \delta_{syn}$ would be as closely aligned with representativeness as the treatment that was designed to change it.
Both anchors represent strong effects.
If even a confounder that matches both anchors simultaneously does not render the effect compatible with zero, we consider the answer to RQ1 reliable.
Since the bias formula is exact for linear models, its application on the latent scale of the probit model is an approximation~\cite{mood2010logistic}.
We therefore interpret the analysis as an assessment of robustness rather than as an exact bound.

\section{Study Results}

\subsection{Participant Information}

\begin{figure*}
    \centering
    \subfloat[Years of experience.]{%
        \includegraphics[width=0.25\textwidth]{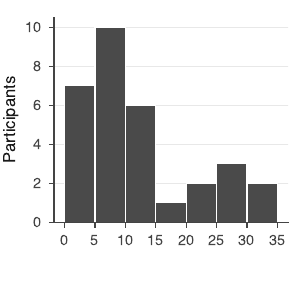}%
        \label{fig:participants-years}}
    \hfil
    \subfloat[Requirements-based activities.]{%
        \includegraphics[width=0.25\textwidth]{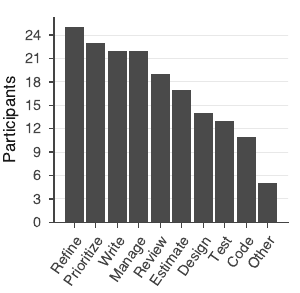}%
        \label{fig:participants-activities}}
    \hfil
    \subfloat[Requirements frequency.]{%
        \includegraphics[width=0.25\textwidth]{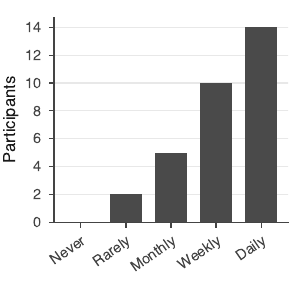}%
        \label{fig:participants-frequency}}
    \hfil
    \subfloat[Professional background.]{%
        \includegraphics[width=0.25\textwidth]{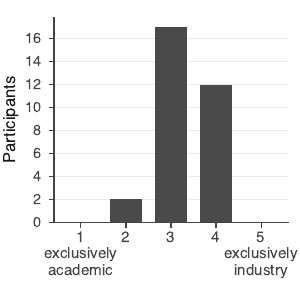}%
        \label{fig:participants-background}}
    \caption{Participant demographics ($N=31$).}
    \label{fig:participants}
\end{figure*}

From a total of 54 responses, 17 participants dropped out on page~1 (context elicitation), two dropped out on page~2 (task description) and four dropped out on page~3 (main rating task).
This left 31 complete responses, none of which met our exclusion criteria of never working with requirements or having a purely academic background.

Our sample spans a diverse range of professional experience, with a median of 7 years in software engineering (Figure~\ref{fig:participants-years}).
The reported domains cover e-commerce, transportation, medical technology, automotive, and several others.
Fourteen participants selected specification-style requirements as more representative of their work, while 17 selected the user story format.
Most participants write, refine, and prioritize requirements (Figure~\ref{fig:participants-activities}).
Moreover, most work with requirements regularly, with only two reporting rare engagement (Figure~\ref{fig:participants-frequency}).
Consistent with our recruitment goal, the professional background leans toward industry (Figure~\ref{fig:participants-background}).

\subsection{Requirements Analysis}

\begin{figure}
    \centering
    \subfloat[Parse-tree depth.]{%
        \includegraphics[width=0.46\columnwidth]{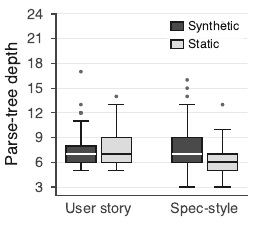}%
        \label{fig:complexity-depth}}
    \hfil
    \subfloat[Sentence Length.]{%
        \includegraphics[width=0.46\columnwidth]{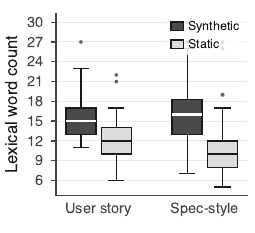}%
        \label{fig:complexity-lexical}}
    \caption{Complexity of synthetic vs.\ static requirements used in this study (synthetic user stories: $n=170$; synthetic spec-style: $n=140$; static user stories: $n=50$; static spec-style: $n=50$).}
    \label{fig:complexity}
\end{figure}

To gauge the complexity of the synthetic requirements, we compare them to the static requirements used in this study.
Synthetic requirements show similar parse-tree depth to the static requirements in our sample for both the user story and the specification-style format (Figure~\ref{fig:complexity-depth}).
Sentence length, by contrast, differs more visibly.
Synthetic requirements tend to be longer than their static counterparts used in this study in both formats (Figure~\ref{fig:complexity-lexical}).

Two authors independently vetted all synthetic-smelly requirements to check whether the smell mutation step injected the smells correctly.
Both authors unanimously agree that all smells have been correctly included with respect to the five intended types (perfect agreement).

\subsection{Results from the Bayesian Data Analysis}

\begin{figure}
    \centering
    \includegraphics[width=1\linewidth]{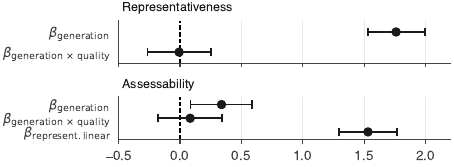}
    \caption{Posterior distribution of the RQ-relevant coefficients of the \textit{representativeness} model (top) and \textit{assessability} model (bottom).}
    \label{fig:posteriors}
\end{figure}

\textit{Representativeness affects Assessability (RQ1).}
Figure~\ref{fig:posteriors} reports the posterior estimates and 95\% CIs of the coefficients on which the five RQs hinge.
Our \texttt{assessability} model estimates the effect of perceived \texttt{representativeness} on \texttt{assessability} (bottom whisker in Fig.~\ref{fig:posteriors}).
Because \texttt{representativeness} enters the regression as a five-level ordinal predictor, its effect is decomposed by default into four orthogonal polynomial contrasts (linear, quadratic, cubic, and quartic), which together exhaust any function of the five levels.
The linear contrast $\beta_{represent.linear}$ has a 95\% CI entirely above zero (Figure~\ref{fig:posteriors}), indicating a strong positive effect.
The three higher-order contrasts (quadratic, cubic, and quartic) are more difficult to interpret, thus we excluded them from Figure~\ref{fig:posteriors} (full posteriors in our online material\textsuperscript{1}).
Nonetheless, the cubic contrast also has a 95\% CI entirely above zero, indicating that the effect of representativeness on assessability exhibits some curvature.
Since \texttt{generation}, a confounder of the effect of \texttt{representativeness} on \texttt{assessability}, is controlled for in the model, this coefficient admits a causal interpretation under the assumptions specified in Figure~\ref{fig:dag}: requirements that practitioners rated as more representative of their work were also rated as substantially better assessable.

\begin{figure}
    \centering
    \includegraphics[width=0.9\linewidth]{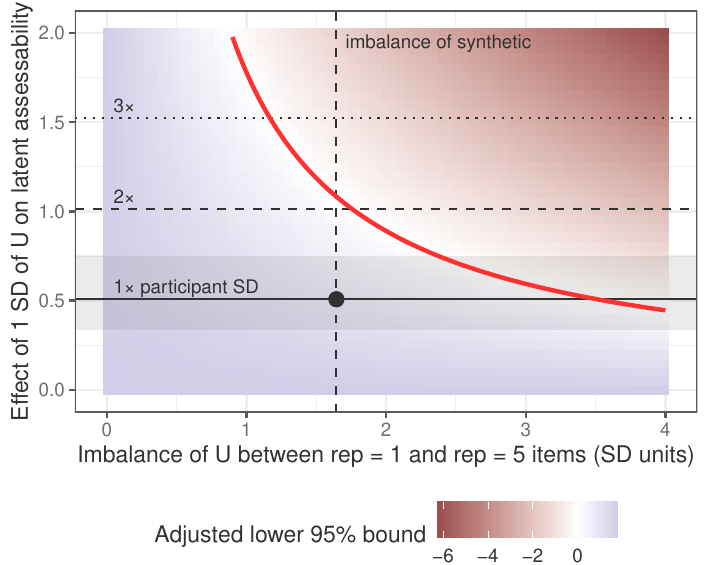}
    \caption{Result of the sensitivity analysis, showing that the empirical anchor (black point) lies in the region below the red line where the effect survives.}
    \label{fig:sensitivity}
\end{figure}

The sensitivity analysis in \Cref{fig:sensitivity} shows that a confounder $U$ with $\gamma = \sigma \approx 0.50$, i.e., one that affects assessability as strongly as the entire variation between participants, would require $\delta = 3.61$ to render $\Delta$ compatible with zero.
This is about 2.2 times the imbalance $\delta_{syn} = 1.64$ of the experimentally manipulated factor \emph{generation}.
Conversely, a confounder with $\delta = \delta_{syn}$ would require $\gamma \approx 1.1$, about 2.2 times $\sigma$.
We consider a confounder of this strength implausible, though the effect of representativeness on assessability is supported by this sensitivity analysis rather than identified by randomization.

\textit{Synthetic Requirements (RQ2 \& RQ3).}
For both representativeness and assessability, the credible interval of $\beta_{generation}$ (first and third rows in Fig.~\ref{fig:posteriors}) lies entirely above zero.
We therefore conclude with at least 97.5\% posterior probability that synthetic study objects increase both the perceived representativeness (RQ2) and the perceived assessability (RQ3) of requirements relative to static study objects sampled from public datasets.
The marginal effects of \texttt{generation} on both outcomes (figures in our online material\textsuperscript{1}) illustrate the size of this shift.
For representativeness, the most likely response was \textit{strongly disagree} in the static condition (about 36\% likelihood) and \textit{strongly agree} in the synthetic condition (about 57\% likelihood), indicating a strong effect of \texttt{generation}.
For assessability, the effect is much weaker but points in the same direction.
The reason the direct effect of \texttt{generation} on \texttt{assessability} is smaller than that on \texttt{representativeness} is the strong indirect path identified for RQ1: most of the assessability gain flows through representativeness.

\textit{Smell Mutation (RQ4 \& RQ5).}
To assess whether LLM-based smell mutation alters the effect of generation approach, we inspect the interaction coefficients $\beta_{generation \times quality}$ of the representativeness and assessability models, respectively (second and fourth rows in Fig.~\ref{fig:posteriors}).
For both outcomes, the 95\% CI includes zero near the center of the posterior distribution.
Therefore, any assumed moderation of the \texttt{generation} effects by requirement \texttt{quality} could be almost as plausibly negative as positive.
Yet either way, the strength of the moderation tends to be small.

\section{Discussion}

\subsection{Threats to Validity}

We discuss the threats to the validity of our study along the four
established validity types~\cite{wohlinExperimentationSoftwareEngineering2024}.
Some threats are inherent to the proposed approach of generating study objects with LLMs, while others are specific to the present study.

\textit{Internal validity.}
By design, the study objects are tailored to each participant's context and therefore differ in content from one participant to the next.
Participant and stimulus set are thus partly confounded, which is a degree of experimental control we deliberately trade for representativeness.
One concrete consequence is that thin or ambiguous context descriptions can affect the generated requirements and, in turn, the measured variables.
A few participants indeed provided only sparse descriptions, for instance, giving ``complete vehicle'' as the system, or describing more than one system.
Such inputs may have constrained the relevance of the generated requirements; even so, we still observed a clear difference in perceived representativeness between the static and synthetic conditions.
A further concern is blinding: because the synthetic study objects are personalized, it is likely that participants recognized some requirements as LLM-generated; some participants also explicitly noted this in the free-text feedback.
We do not know how this knowledge could have influenced the responses.
Awareness about the LLM generation could have positively impacted the outcome variables if participants wanted to help us authors achieve ``good results'', i.e., social-desirability bias~\cite{wohlinExperimentationSoftwareEngineering2024}.
We tried to mitigate this risk by reminding participants on Page~2 of the survey to respond honestly, that there are no right or wrong answers, and that they should answer with what feels right to them.
On the other hand, awareness about the LLM generation could have negatively impacted the outcome variables if participants are generally skeptical or distrusting in LLM-generated content.
To further mitigate this risk, participants were not told that any requirements were static, and the presentation order was randomized.
The same reasoning applies to the blinding between clean and smelly requirements, which apparently, is less likely to have leaked.

\textit{External validity.}
Our findings rest on a sample recruited through personal invitations and thus is not a random sample from the population of software engineering practitioners.
In SE research, a truly random sample is usually not possible to achieve.
In addition, the synthetic requirements were produced with a single LLM, and results may differ for other LLMs.

\textit{Construct validity.}
We operationalized representativeness and assessability each through a single item, although both are multifaceted: representativeness can concern content or style, and assessability can be judged with a different task in mind (writing, reviewing, coding, or testing).
We deliberately elicit one holistic judgment per construct rather than querying each facet separately, which could cognitively overload participants and dilute the overall gut feeling that ultimately governs whether they can engage with a study object.

Moreover, both operationalizations are self-reported.
For representativeness, we see this as a necessity and a first-person judgment as the most reasonable operationalization.
Assessability, in contrast, is a proxy: it reflects participants' reported confidence to act, not how soundly they would actually act.
By design, we study this proxy first.
Confidence is the more immediate question: for representativeness to affect how soundly practitioners act, it must first affect how confidently they act.
Measuring behavior directly would instead require fixing a single task and object type which would also entangle the representativeness effect with task design and individual skill.
Moreover, such a design assumes an agreed standard of correct action (i.e., ground-truth), which, in the case of context-dependent phenomena, may not be available (e.g., requirements quality).
Nonetheless, in future work, task-based designs (see Section~\ref{subsec:future-work}) are necessary to support the results from our judgment study with behavioral measures.

In addition, the static requirements were already matched to each participant's preferred requirement format, so format adaptation is present in \emph{both} conditions.
The synthetic condition adds only the application domain and system description on top, meaning that any observed benefit of context adaptation is therefore a conservative estimate and would likely be larger if requirement format were varied as well.

\textit{Conclusion validity.}
Due to the non-determinism of LLMs, the synthetic requirements are not perfectly reproducible, which concerns the reliability of the treatment implementation: even an identical context would not yield the exact same requirements.
We accept this variability as an unavoidable property of the proposed approach.
To support comparability and replication across studies, we report our LLM use following the reporting guidelines of Korn et al.~\cite{korn2026reporting}.

\subsection{General Observations}

We observe that the representativeness of requirements has a strong positive effect on participants' perceived confidence to act on the material (RQ1).
Thereby, this study provides direct empirical evidence about the assumed link between representativeness and assessability in the case of requirements.
We hypothesize that the same holds for other object types that are relevant to SE research, such as code and models.
This draws the external validity of empirical studies, which rarely consider the context-fit of their study objects, into question.

We also investigated whether LLM-generated requirements are suitable as context-adapted study objects in empirical research, and whether they can additionally be mutated to carry requirements smells. 
Our controlled experiment yields positive evidence on both.
Participants perceived the synthetic requirements as more representative of their own work and as better assessable for their typical tasks than the static requirements sampled from public datasets (RQ2, RQ3).
The assessability gain stems primarily from improved representativeness, with a smaller direct effect of generation mode, plausibly reflecting that LLM generated requirements carry an inherent property that makes them more assessable (e.g., better grammar or lack of foreign domain terms).
Injecting artificial smells did not measurably impact either perceived representativeness or assessability (RQ4, RQ5).

The free-text feedback reveals that several participants were positive about the contextual fit.
One respondent stated: ``I was highly impressed by how well the (supposedly) generated requirements matching my application domain present real world use cases.''
A recurring critique, however, was that the synthetic requirements came across as too high-level.
One participant described them as ``too complex and partly too nonspecific,'' and
another noted that some ``lacked enough context for implementation or
testing''.
While some participants recognized that requirements were LLM-generated, none indicated awareness that others were static; hence, we cannot attribute individual comments to the synthetic or the static condition specifically.

Looking at the logs, we see that both the requirements generation and requirements mutation step have been completed for every participant leaving Page~1.
Hence, we conclude that no participant dropped out during the two waiting screens, which combined lasted approximately 15~seconds.
For the just-in-time generation approach to be useful, it is important that the waiting time does not lead to a higher drop out rate.

Regarding the smell mutation step, we observe that all smells have been correctly generated and all represent plausible, naturally-sounding cases.
However, we also note some homogeneity in the generated smells.
This is particularly the case for the weak verb smell, for which the LLM most often injected the verb ``may''.
While these examples are correct smell representations, they demonstrate little diversity.
We assume the reason could be that the LLM adhered closely to the examples provided in the prompt.

\subsection{Relation to Existing Evidence}

Our observations are consistent with the broader evidence on LLM-generated requirements.
Like prior studies reporting high consistency and alignment of LLM-generated requirements~\cite{krishnaUsingLLMsSoftware2024a,
quattrocchiCanLLMsGenerate2026}, we find the synthetic requirements to be stylistically consistent and well aligned with the systems our participants described.
A more notable contrast concerns complexity.
In a related study, synthetic requirements were found to be syntactically simpler than those in public datasets~\cite{femmerDescriptionComparativeAnalysis2025a}.
We find no such under-complexity here: the synthetic requirements are comparable in parse-tree
depth to the static requirements and even tend to be longer (Figure~\ref{fig:complexity}), also compared to the other datasets investigated in the previous study.
Two factors plausibly explain this shift.
First, the previous work already showed that supplying a detailed project and domain description yields more realistic complexity, which is exactly the kind of input our context elicitation collects. 
Second, gpt-5.5 is a more capable model than the gpt-4.1 used previously.
While we cannot disentangle the two, both suggest that the under-complexity reported earlier should not be treated as an inherent property of LLM-generated requirements.

\subsection{Recommendations for Research}

Our approach of \textit{context-adapted study objects} raises assessability through increasing representativeness.
If researchers find other means of supplying context-adapted objects that make them more representative of the participants' working context (e.g., customization ``by hand''), this could likewise raise assessability.
Herein, we propose the use of our approach as \textit{one way} of obtaining more representative objects, and thereby, reducing the threat to external validity from \textit{setting}.

The proposed approach is not specific to requirements, and of course, not specific to requirements smells.
In principle, the same scheme---eliciting a short participant-supplied working context, conditioning an LLM on it, and using the output as a study stimulus---could be instantiated for other SE study objects.
In fact, LLMs are already heavily used to generate several SE artifacts in practice, including source code, unit tests, and code documentation~\cite{hou2024large}.
Whether such instantiations of the approach would raise perceived representativeness, and whether the representativeness-to-assessability link we observe here would carry over, are open empirical investigations that the present study does not answer.
Nonetheless, we have several recommendations for researchers that plan to employ the approach, for which we conjecture that they are valid for other types of study objects as well.

\textit{When to apply the approach.}
Adopting context-adapted study objects is a deliberate methodological tradeoff.
Fixed study objects are perfectly uniform across participants, which simplifies treatment standardization and allows per-item comparisons across participants.
Context-adapted study objects vary across participants by design, but each participant sees material grounded in their own working context, supporting external validity in the setting dimension as discussed in Section~\ref{subsec:validity}.
We therefore recommend the approach primarily for context-dependent research questions, where representativeness is particularly important and some loss of stimulus control is acceptable.
For studies that depend on holding stimuli perfectly constant, fixed objects remain preferable.
Within this scope, the approach applies in principle to human-subject studies regardless of participant background (students or practitioners), provided a meaningful working context can be elicited.
We are more cautious about non-human study subjects, specifically LLMs, due to information leakage and self-preference bias risks (e.g., first injecting and then detecting requirements smells with the same LLM).

\textit{How to apply the approach.}
We recommend vetting the generation approach both before and after data collection.
A pre-hoc check confirms that the generated material carries the intended properties.
For research that requires stimuli (e.g., smells) to be more diverse, we assume more variability could be reached by having more diverse examples, different prompt variants, or inviting the LLM to be creative.
Other research might require a higher level of control on the generated stimuli, favoring strict prompt-adherence.
A post-hoc check, similar to our re-inspection of the mutated requirements, then verifies validity on the material participants actually saw.
Such a post-hoc check, in consequence, has impacts on the scalability of the approach.
Where the design permits, we further recommend eliciting participants' own representativeness judgments, since they surface mismatches that researchers may not anticipate.
Finally, the context-elicitation instrument deserves the same care as the generation prompt, as thin or ambiguous descriptions might degrade the result.
Prompting participants for sufficient detail, e.g., through examples or guided fields, and piloting the instrument are worthwhile.

\textit{Reporting and ethics.}
Although LLM non-determinism precludes exact reproduction, studies should report their LLM use.
We recommend following prompting reporting guidelines, such as the one of Korn et al.~\cite{korn2026reporting} to which we adhered.
Finally, while the approach deliberately avoids requesting confidential material, the elicited context descriptions can themselves be sensitive and need to be handled with care.

\subsection{Future Work}
\label{subsec:future-work}

Our study examined a single object type, requirements, and measured assessability through self-reports.
Two extensions are therefore particularly important.
First and foremost, the study should be replicated with other study objects, such as source code, tests, or architectural diagrams, to test whether the representativeness–assessability link we observe generalizes beyond requirements, as we conjecture.
Second, self-reported assessability should be complemented with revealed behavior, by having participants act upon the study objects through specific tasks.
Such task-based designs would also allow investigating whether more realistic settings raise participants' engagement and motivation to take a study seriously, as suggested by Sj{\o}berg et al.~\cite{sjobergConductingRealisticExperiments2002}.

\section{Conclusion}

External validity in empirical SE has long been debated in terms of who participates in our studies, and much less in terms of the study objects participants work on.
In our experiment with 31 practitioners, the perceived representativeness of requirements had a strong positive effect on their assessability, i.e., participants' confidence in acting on them.
This provides support for a threat that Wohlin et al.~\cite{wohlinExperimentationSoftwareEngineering2024} already list in their catalog of external validity, but for which empirical evidence was missing.
We also proposed generating
context-adapted study objects just-in-time from a short description of
the participant's working context.
Instantiated for requirements, these were perceived as both more representative and more assessable than objects sampled from public datasets, primarily through the representativeness gain.
%This is not to say that synthetic requirements are more \textit{real} than static requirements sampled from public datasets.
%In fact, the static requirements we used in the study were real and synthetic requirements by definition are not real.
%We do propose, however, that context-adapted synthetic requirements allow for producing more \textit{realistic} settings as they better represent the practitioners' domain and expertise.
We therefore recommend that researchers consider study-object representativeness when designing experiments, and see context-adapted objects as one way to achieve it.
Whether the representativeness-assessability link carries over to other object types and downstream behavioral effects is an open question and we hope this first step invites the community to pursue it.

\bibliographystyle{IEEEtran}
\bibliography{references}

\end{document}